# Mid-infrared photothermal single-live-cell imaging beyond video rate

Genki Ishigane[1,+], Keiichiro Toda[1,+], Miu Tamamitsu[1,2], Hiroyuki Shimada[2], Venkata Ramaiah Badarla[2], and Takuro Ideguchi[1,2,*]

[1] Department of Physics, The University of Tokyo, Tokyo, Japan

[2] Institute for Photon Science and Technology, The University of Tokyo, Tokyo, Japan

[+] These authors contributed equally to this work

[*] Corresponding author: ideguchi@ipst.s.u-tokyo.ac.jp

**Advancement in mid-infrared (MIR) technology has led to promising biomedical applications of MIR spectroscopy, such as liquid biopsy or breath diagnosis. On the contrary, MIR microscopy has been rarely used for live biological samples in an aqueous environment due to the lack of spatial resolution and the large water absorption background. Recently, mid-infrared photothermal (MIP) imaging has proven to be applicable to 2D and 3D single-cell imaging with high spatial resolution inherited from visible light. However, the maximum measurement rate has been limited to several frames/s, limiting its range of use. Here, we develop a significantly-improved wide-field MIP quantitative phase microscope with two orders-of-magnitude higher signal-to-noise ratio than previous MIP imaging techniques and demonstrate single-live-cell imaging beyond video rate. We first derive optimal system design by numerically simulating thermal conduction following the photothermal effect. Then, we develop the designed system with a homemade nanosecond MIR optical parametric oscillator and a high full-well-capacity image sensor. Our high-speed and high-spatial-resolution MIR microscope has great potential to become a new tool for life science, in particular for single-live-cell analysis.**

Vibrational imaging such as Raman scattering and mid-infrared (MIR) absorption imaging has attracted attention in life science[1,2], e.g., in the field of single-cell biology, because its label-free capability can solve the problems associated with fluorescence imaging, such as cell damage or death due to cytotoxicity, difficulty in continuous and quantitative measurements due to photobleaching, and undesired functional modification of the labeled intracellular biomolecules[3,4]. Most single-cell vibrational imaging techniques exploit Raman scattering, and the state-of-the-art coherent Raman scattering (CRS) microscopes have achieved high-speed imaging at video rates[5,6]. These high-speed CRS imaging systems have made significant impacts in the field of vibrational imaging and triggered drastic expansion of the related research, including instrumental developments[7,8,9,10,11] and biological applications[12,13], particularly single-cell analysis. On the other hand, MIR absorption imaging is rarely used for detailed observation of single cells because of the low spatial resolution of 2-10 μm restricted by the diffraction limit of MIR light and the strong background absorption by the water surrounding the cells. However, MIR absorption imaging has great potential for life science due to the large absorption cross-section of the MIR absorption process (~$10^8$ times higher than that of Raman scattering) with low photo-damage to biomolecules. CRS imaging, in contrast, exploits tightly-focused ultrashort laser pulses to trigger nonlinear optical effects to perform sensitive measurements, causing undesired multiphoton electronic transitions that can cause deleterious effects to live cells[14]. MIR absorption avoids this problem because it is a single-photon linear absorption process with low photon energy. In addition, imaging with MIR light holds a possibility of obtaining information that Raman spectra have not revealed until now. MIR spectra provide rich information on biomolecules that dominantly exist in a cell, such as proteins and water, more specifically, e.g., the secondary structure of proteins via the amide band[15,16] and the state of water molecules via the OH band[17].

MIR photothermal (MIP) imaging is an emerging technique that has been studied in the last several years[18,19,20,21,22,23,24,25,26], which can solve the above-mentioned problems and enables single-cell MIR absorption imaging. In this technique, MIR molecular absorption induces local heating in the sample, and the resulting change in refractive index is detected as changes in optical parameters such as phase[20,21,22,24], reflectance[19,23], or scattering intensity[18,25] of visible probe light. Hence, one can obtain information on the spatial distribution of MIR absorption with a sub-μm resolution. Moreover, as long as the MIR light reaches the cell, it is possible to capture the intracellular change in refractive index due to the transparency of visible light in biological samples, even if the MIR light is subsequently absorbed by the water behind the cell.

However, the performance, particularly the frame rate, of current MIP imaging systems has not yet reached the level of the state-of-the-art CRS imaging systems. MIP imaging techniques can be classified into "point-scanning" or "wide-field" configurations. The pioneering work on single-cell imaging was demonstrated based on the point-scanning configuration[18], in which MIR and visible light emitted from a pulsed quantum cascade laser (QCL) and a CW laser diode, respectively, were coaxially focused on a sample. In this configuration, images were taken by scanning the sample stage, and the maximum frame rate was limited to ~0.1 Hz for taking 100 pixels x 100 pixels due to the low scanning speed of the stage and low detection efficiency of the photothermal signals. Wide-field configurations have solved this problem[19,20,21,22,23,24,25], in which the entire field of view (FOV) is irradiated with MIR and visible light, and the wide-field photothermal signals in the FOV are detected at once with a CMOS image sensor. Since the maximum image-acquisition rate is determined as half the frame rate of the image sensor, molecular vibrational imaging beyond video rate can be realized if a high SNR is achieved. However, the image-acquisition rate in cell measurement with sub-μm spatial resolution remains in the range of 0.1 to 2 Hz[19,21,24,25] for the state-of-the-art wide-field systems because of their low SNR due to the following reasons: (1) the low photothermal signal owing to the decrease in MIR fluence caused by wide-field illumination and the signal saturation caused by thermal diffusion, and (2) the low detection sensitivity of wide-field microscopes limited by optical shot noise due to the use of CMOS image sensors with low full-well capacity. Supplementary Note 1 summarizes these parameters of existing wide-field systems.

In this work, we develop a high-SNR MIP imaging system with a high-intensity MIR nanosecond optical parametric oscillator (OPO) and highly sensitive quantitative phase imaging (QPI) using a high full-well-capacity CMOS image sensor, with which we, for the first time, realize live-cell MIP imaging beyond video rate. First, we perform thermal conduction simulations to derive the optimal pulse duration and repetition rate of MIR and visible light for wide-field MIP imaging: less than 10 ns and ~1 kHz, respectively. Then, we develop a wavelength-tunable MIR nanosecond OPO with a periodically poled lithium niobate (PPLN) crystal that meets these requirements with ~10-μJ pulses (~100-times higher pulse energy compared to that of a QCL in our previous study[22]) in the wavenumber region of 2,600-3,450 cm$^{-1}$. Next, we employ a high-full-well capacity CMOS image sensor for quantitative phase measurements, capturing ~100-times more photons than conventional image sensors. The SNR of our system is evaluated to be ~210-times higher than that of the previous work[21] with the highest SNR amongst the wide-field configurations, to the best of our knowledge (see Discussion for details). With the developed system, we perform MIP imaging of live COS7 cells in the 2,925-cm$^{-1}$ band with a high SNR of 89 at a record rate of 50 fps. The high-SNR and high-speed capabilities of our microscope are expected to be beneficial in video-rate observation of intracellular dynamics and for high-speed broadband MIR spectral image acquisition over several hundred cm$^{-1}$ in less than 1s.

**Results**

**Derivation of the optimal pulse duration and repetition rate of MIR and visible light by thermal conduction simulations.** We consider the optimal pulse duration and repetition rate of MIR and visible light for wide-field MIP imaging by exploiting thermal conduction simulations. Thermal diffusion causes degradation of spatial resolution and saturation/decay of the amount of signals in the MIP imaging. The change in optical phase-delay of visible light due to the local temperature rise, which we call the MIP phase change, is expressed as

$$\Delta\theta(x,y,t) \sim \frac{2\pi}{\lambda}\frac{dn}{dT} \int \Delta T(x,y,z,t)dz, \qquad (\text{Eq.1})$$

where x, y, z denote spatial coordinates, $t$ the time, $\lambda$ the wavelength of visible light, $dn/dT$ the thermo-optic coefficient of the sample, and $\Delta T$ the local temperature change. The temporal evolution of the MIP phase change under various MIR excitation conditions in an aqueous environment is calculated by solving the 3D heat conduction equation,

$$\frac{\partial \Delta T(x,y,z,t)}{\partial t} = \left(\frac{\partial^2}{\partial x^2} + \frac{\partial^2}{\partial y^2} + \frac{\partial^2}{\partial z^2}\right) \nu \Delta T(x,y,z) + \frac{I(x,y,z,t)\alpha(x,y,z)}{\rho c_{\mathrm{p}}}, \quad (\text{Eq.2})$$

where $\nu$ denotes the thermal diffusivity, $I$ the pulse fluence per unit time, $\alpha$ the absorbance, $\rho$ the density, $c_{\mathrm{p}}$ the specific heat capacity.

To derive the optimal pulse duration of MIR light, we calculate the spread of the spatial profile (Fig. 1a) and the phase change (Fig. 1b) in the MIP phase change image with respect to the pulse duration of MIR light. We assume the initial heat spots (target objects) are spheres with a diameter of 500 nm, 2 µm, and 10 µm in aqueous environments. They have the same thermal diffusivity as water and are continuously heated during irradiation with the MIR pulse. We assume that the visible probe pulse is sufficiently shorter than the MIR pulse and illuminated at the end of the MIR pulse. Figure 1e shows the timing chart of the MIR and visible pulses. The results show that a longer pulse builds up the MIP phase change, but too much elongation leads to degradation of the spatial resolution and signal saturation due to heat diffusion, particularly for small objects with a large surface/volume ratio. For example, when observing the 500-nm object, a 100-ns MIR pulse deteriorates the spatial resolution by a factor of 1.3 (Fig. 1a). Some works exploit a CW MIR light source[20,23], and in such cases, the thermal spread is ~4.8-times larger than the actual size of the target. Meanwhile, the MIP phase change of the 500-nm object is saturated after illuminating for 100 ns, causing a lack of quantitative capability with longer MIR pulses (Fig. 1b). These results show that MIR pulses of ~10 ns or shorter are desirable for quantitative imaging by confining the generated heat within a near-diffraction-limited spot of visible light.

Next, to derive the optimal pulse duration and delay of the visible light, we calculate the spread of the spatial profile (Fig. 1c) and the phase change (Fig. 1d) in the MIP phase change image with respect to the delay of the visible probe light from the end of the MIR excitation. In this calculation, the MIR pulse duration is 10 ns, and the visible probe pulse is sufficiently shorter than the MIR pulse. The results show that the probe delays longer than 10 ns cause degrading the spatial resolution (Fig. 1c) and decreasing the phase change (Fig. 1d) of the MIP phase change image. For example, when observing the 500-nm object with a delay of 100 ns, the radius of the MIP phase change becomes 1.9-times larger than the object, and the MIP phase change decays to 50% of that with a delay of 0 s. This simulation shows that it is desirable for visible probe pulses to have a similar or shorter pulse duration than MIR pulses, i.e., < 10 ns, with the delay time shorter than the pulse duration.

Finally, to derive the optimal pulse repetition rate, we calculate the thermal diffusion time of a heated spot with an FWHM diameter of 91 µm in an aqueous environment which reflects the condition of our following experiment (Fig. 1f). Note that, when measuring cells, the thermal diffusion time over the entire FOV does not depend on the size of the target objects but on the spot size of MIR light due to water absorption. To avoid potential thermal damage to samples due to a thermal pile-up, the induced photothermal heat should be diffused off when the next MIR pulse arrives at the sample. The result shows that this condition is sufficiently achieved at 1 ms after the arrival of the first MIR pulse. Hence, it is desirable for wide-field single-cell imaging to exploit a pulse repetition rate of ~1 kHz.

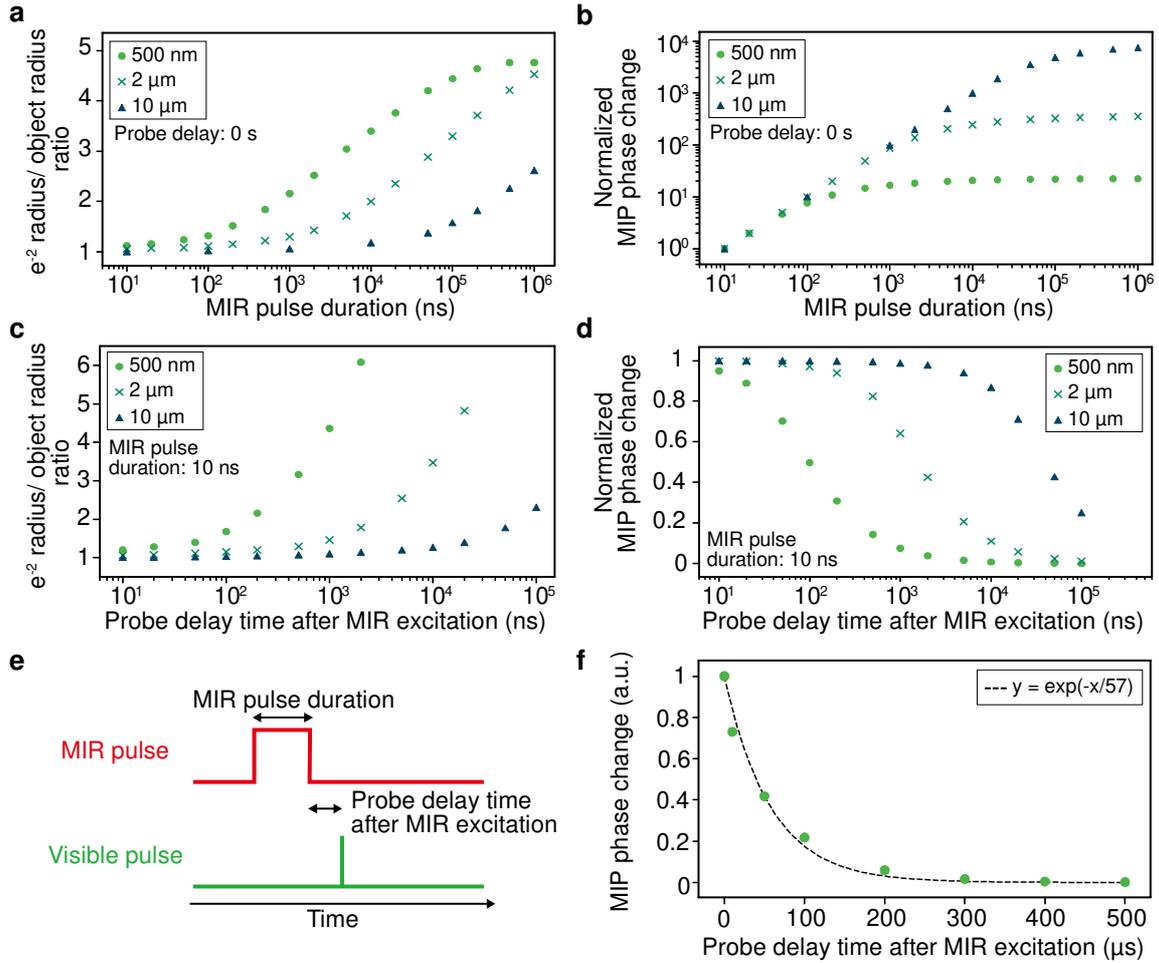

**Figure 1 Derivation of optimal pulse duration and repetition rate of MIR and visible light by thermal conduction simulations. a** Degradation of the spatial resolution in MIP imaging originating from thermal diffusion depending on the MIR pulse duration. The horizontal axis is the MIR pulse duration, and the vertical axis is the ratio of the $e^{-2}$ radius in the MIP phase change image and the radius of the original spheres. **b** Saturation of the MIP phase change depending on the MIR pulse duration. MIP phase changes at the centers of the spheres are plotted against MIR pulse durations, which are normalized by that with the MIR pulse duration of 10 ns. The probe delay time after MIR excitation is 0 s for **a** and **b**. **c** Degradation of the spatial resolution in MIP imaging originating from thermal diffusion depending on the probe delay time after MIR excitation. **d** Decay of the MIP phase change depending on the probe delay time after MIR excitation. The MIP phase changes are normalized by that with the probe delay of 0 s. MIR pulse duration is set to 10 ns for **c** and **d**. **e** Pulse duration and timing chart of the MIR and visible pulses. **f** Temporal decay of the MIP phase change for water (10 µm thickness) sandwiched between two $CaF_2$ substrates. The vertical axis shows the MIP phase change at the center of the heated spot. The pulse duration of the MIR light is assumed to be much shorter than the thermal decay time. The spatial distribution of the MIP phase change is assumed as a gaussian function (FWHM = 91 µm) along the x- and y-axes, determined by the intensity profile of the MIR spot, and an exponential function along the z-axis that decays after 16 µm, which is derived from the Lambert-Beer law.

**High-SNR MIP-QPI system.** The principle and the schematic of high-SNR MIP-QPI are shown in Fig. 2. MIR light with a narrow spectral width at a certain wavenumber is irradiated widely over the sample. Resonant molecules absorb the MIR light and are excited to their vibrational states. The molecular vibrations relax by transferring their energy to the surrounding medium in the form of heat, causing thermal expansion and thus a change in local density. The resulting change in refractive index in the vicinity of the target molecules (i.e., MIP effect) is captured as a change in optical phase delay of the transmitted visible light in the QPI system. As shown in Fig. 2a, a MIP image is generated by taking the difference between the phase images captured in the MIR-ON and -OFF states. Amongst the available detection methods of the MIP effect, QPI is the optimal method for the quantitative measurement of intracellular molecular distributions. For example, phase-contrast microscopy[27] suffers from image artifacts such as halos, while dark-field[28] and interferometric scattering (iSCAT)[29] microscopes sacrifice a part of the spatial-frequency information. QPI does not have those drawbacks and provides quantitative MIP images. In addition, detailed morphology with dry-mass information can also be obtained from the quantitative phase image in the MIR-OFF state. Therefore, one can make a correlation analysis between the spatial distribution of target molecules and cell organelles.

Figure 2b shows the schematic of our high-SNR MIP-QPI system developed in this study. Two Nd:YAG Q-switched lasers (1,064-nm wavelength, 1-kHz repetition rate, 6-ns pulse duration) (NL204, Ekspla) are used to generate visible and MIR pulses via nonlinear wavelength conversions. The visible light pulses (532-nm wavelength, 1-kHz repetition rate, 5-ns pulse duration) are provided by second harmonic generation (SHG) with a 15-mm-long LBO crystal. The MIR light pulses (2,800-3,250 cm$^{-1}$ wavenumber tunable, 1-kHz repetition rate, 9-ns pulse duration, ~10-µJ pulse energy) are obtained as idler pulses of a homemade high-intensity nanosecond OPO with a fan-out PPLN crystal[30] (HC Photonics Corporation). The visible pulse is electronically synchronized with the MIR pulse using a function generator. In our current system, we set a delay of ~100 ns between the MIR and visible pulses, which is longer than the theoretical optimum because there is a timing jitter up to 50 ns between the pump and probe pulses. The 100-ns delay guarantees the visible probe pulses come after the MIR pump pulses under this amount of jitter. Note that we can suppress the jitter down to ~1 ns, which is specified in a product specification sheet of our lasers, by appropriate synchronization. The MIR light is intensity-modulated by a mechanical chopper at 50 Hz synchronized with the image sensor frames such that the sensor alternately acquires MIR-ON and -OFF frames (Fig. 2c). The MIR light pulses are loosely focused onto the sample with a spot size of ~80 µm x 80 µm with a ZnSe lens. The visible light pulses from the single-mode fiber are collimated and irradiated onto the sample with a peak fluence of ~30 pJ/µm$^2$ (~400 nJ over 110 µm x 110 µm). Common-path off-axis digital holography is used as QPI[31]. The light transmitted through the sample is replicated by a diffraction grating, and the zeroth-order diffraction light is low-pass filtered with a pinhole placed in the Fourier plane, thus, converted to a quasi-plane wave that acts as the reference light. The first-order diffraction light is used as the object light, which contains information on the optical phase delay induced by the sample. Interference fringes between the two lights are captured as an off-axis hologram with the high-full-well-capacity image sensor (Q-2HFW, Adimec Advanced Image Systems) after relay lenses in 4f configuration, from which the phase image is numerically reconstructed. The spatial resolution of QPI is 440 nm, determined by the NA of the objective lens (LUCPLFLN40X, Olympus).

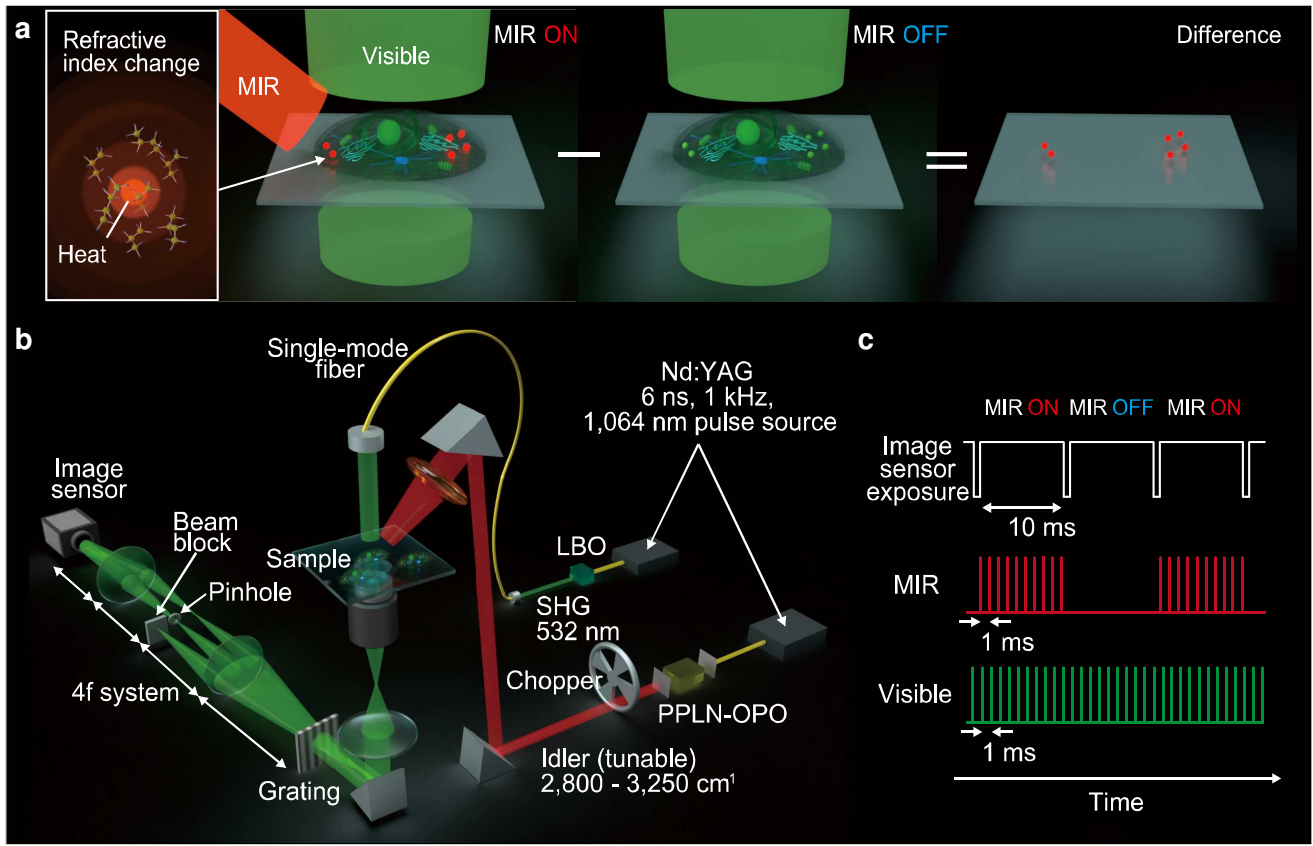

**Figure 2 Principle and schematic of high-SNR MIP-QPI. a** Principle of MIP-QPI. **b** Optical system of high-SNR MIP-QPI. **c** Timing chart of MIR and visible pulses. MIR: mid-infrared, LBO: LiB$_3$O$_5$, SHG: second harmonic generation, PPLN: periodically-poled lithium niobate, OPO: optical parametric oscillator.

**High-intensity nanosecond MIR light source.** We describe the performance of our homemade ns-PPLN-OPO. The crystal is a 50-mm-long fan-out PPLN with a poling period varying from 27.5 to 31.6 µm stabilized at 40 °C. The pulse energy of the pump light from the Nd:YAG Q-switched laser is ~100 µJ. The OPO cavity is resonant with the NIR signal pulses (5,950-6,800 cm$^{-1}$ tunable), and only the MIR idler pulses (2,600-3,450 cm$^{-1}$ tunable) are extracted with a long-pass filter after the cavity. Figure 3a shows the relationship between the MIR wavenumber and the idler pulse energy, which is ~10 µJ between 2,800 and 3,250 cm$^{-1}$. Figure 3b shows the spectrum of MIR light measured by a homemade FTIR spectrometer. The FWHM of the spectrum is ~10 cm$^{-1}$ at each MIR wavenumber, which determines the spectral resolution and is sufficient to resolve absorption peaks of CH$_3$ and CH$_2$ stretching modes (the modes are 20~30 cm$^{-1}$ apart from each other)[32], which are the major signatures of analyzing biological samples in this wavenumber range. Figure 3c shows how the MIP phase change of water changes with respect to the MIR pulse energy under the condition of the MIR wavenumber of 2,918 cm$^{-1}$, the spot size of 69 µm x 69 µm, and the pulse energy ranging from 1.3 to 8.9 µJ. The values of the MIP phase change in the graph are averages of 20 pixels x 20 pixels around the center of the MIR spot. The results show that the MIP phase change increases linearly with the MIR pulse energy over the entire range, which indicates that our OPO yields a ~100-times larger MIP phase change compared to the previous work with a QCL[22]. Finally, the temporal decay of the MIP phase change in water is measured by scanning the time delay between the MIR and visible light pulses (Fig. 3d). One can see that the MIP phase change decays to 1/e at 74 µs and to 3/100 at 500 µs, which is consistent with the results derived from the thermal conduction equation (1/e at 57 µs). Since it is desirable to observe live cells under a similar or larger MIR illumination spot size, a repetition rate of 1 kHz is low enough to avoid potential sample damage induced by a thermal pile-up.

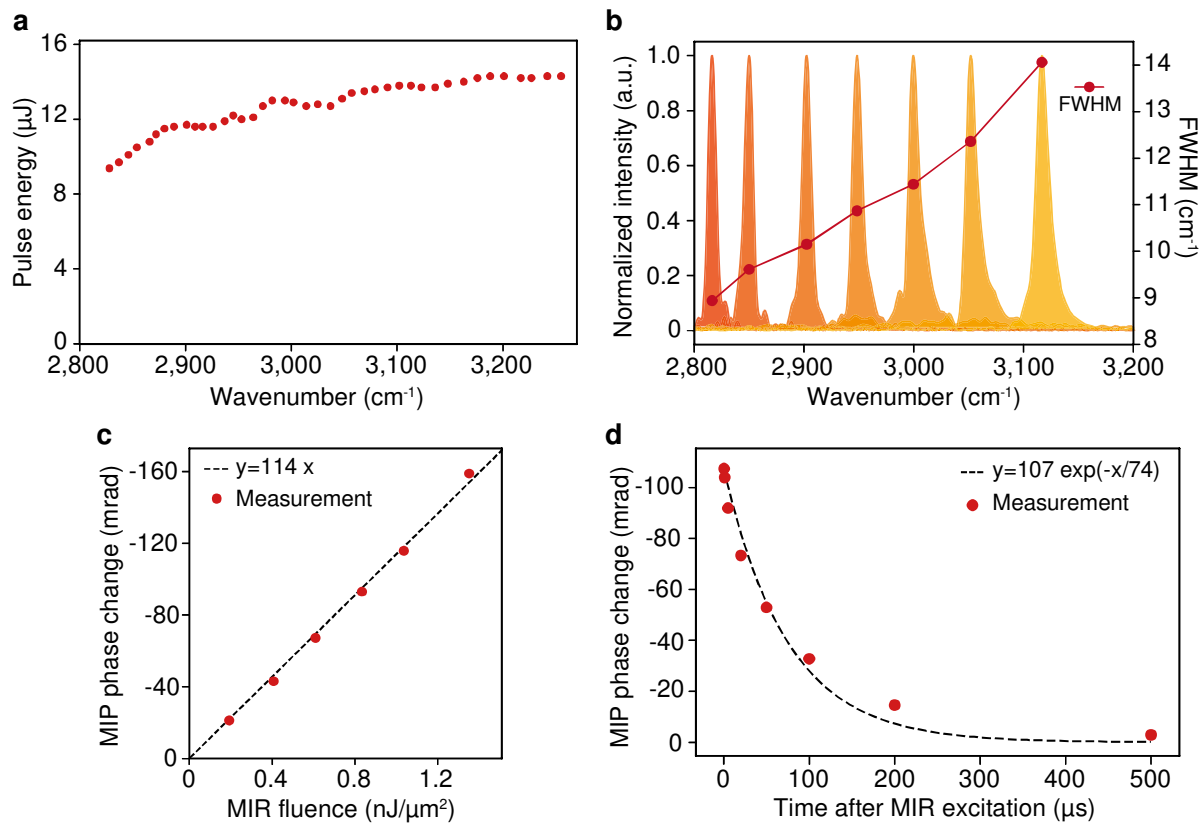

**Figure 3 Basic performances of MIR light source (ns-OPO).** **a** MIR pulse energy in the range of 2,800-3,250 cm$^{-1}$. **b** Spectra of MIR light measured with a homemade FTIR spectrometer. The red points indicate the FWHM of each spectrum. **c** Linearity of MIP phase change with respect to MIR fluence for a water sample excited at 2,918 cm$^{-1}$. **d** Temporal decay of MIP phase change for a water sample with an excitation area of 69 µm x 69 µm. FWHM: full width at half maximum.

**High-precision QPI system.** We discuss noise reduction in phase measurement with QPI by employing a high full-well-capacity image sensor and a high-intensity ns visible light. If the system is mechanically stable enough, the temporal phase noise in QPI can be dominated by optical shot noise. Thus, the precision becomes higher when more light enters the image sensor. The full-well capacity of the image sensor used in our system is 2 Me$^-$/pixel (Q-2HFW, Adimec), which is 200 times larger than that of a conventional CMOS image sensor (10 ke$^-$/pixel, e.g., acA2440-75um, Basler). We perform the following evaluations of the noise reduction.

We examine the dependence of temporal phase noise on the number of electrons contributing to the reconstruction of the phase images (= $N_{electron}$) (Fig. 4a), that is, the average number of electrons in the holograms (see Supplementary Note 2 for the calculation method). The maximum value of $N_{electron}$ is determined as half the full-well capacity of the image sensor. Note that $N_{electron}$ equals the number of incident photons multiplied by the quantum efficiency of the image sensor. We record 100 holograms without a sample and calculate the differences in phase images between adjacent frames. Then, the temporal standard deviation (STD) of the 50 differential images is calculated at each pixel, and the average value of 80 pixels x 80 pixels in the temporal STD map is evaluated as the temporal phase noise. The number of electrons per pixel is estimated from the sensor's digital output value and the full-well capacity. The data points on the left side of the graph are the measurement results using the conventional 10k-e$^-$ image sensor, which is in good agreement with the theoretically estimated phase noise limited by optical shot noise (orange line)[33] (see Methods for details). This indicates that the phase noise can be reduced by detecting more light. The data points on the right are the results using the high-full-well-capacity 2M-e$^-$ image sensor. One can see that the maximum number of detected electrons is ~100-times larger than that measured with the 10k-e$^-$ image sensor, resulting in a reduction of the phase noise by a factor of 7.9 (corresponding to 0.9 mrad with $N_{electron}$= 3.6 x 10$^5$ e$^-$). However, this phase noise is larger than that determined by optical shot noise and is in good agreement with the estimated value (purple line) that includes the effect of sensor noise ($\sigma_{sensor}$ = 572 e$^-$), measured by turning off the laser. It is expected that near optical shot-noise-limited measurement (0.4 mrad of the phase noise) is feasible when the sensor's full-well capacity is used to the maximum extent ($N_{electron}$= 1 x 10$^6$ e$^-$).

Next, we compare the SNR of single-frame MIP phase change images of live cells measured with the two image sensors. Figure 4b shows results for the observation of COS7 cells exploiting MIR light with a wavenumber of 2,975 cm$^{-1}$, a spot size of 80 μm × 80 μm, and pulse energy of 7.1 μJ. The background MIP phase change image of water without cells is subtracted to make the intracellular structures more visible. The spatial STD of 20 pixels x 20 pixels inside the blue box is defined as the phase noise. Note that Fig. 4b has ~√2-times larger noise than the temporal phase noise in Fig. 4a due to the background subtraction process. In the case of the 10k-e$^-$ sensor, the MIP phase change is buried in the phase noise, whereas in the case of the 2M-e$^-$ sensor, intracellular structures such as nucleoli and lipid droplets are clearly seen. The phase noise is 12.2 mrad for the former sensor and 1.6 mrad for the latter (8.6 mrad and 1.1 mrad without water background subtraction, respectively). We verify that the high full-well capacity sensor provides ~7.6-times reduction in phase noise with 85-times higher power of visible light for the live-cell imaging.

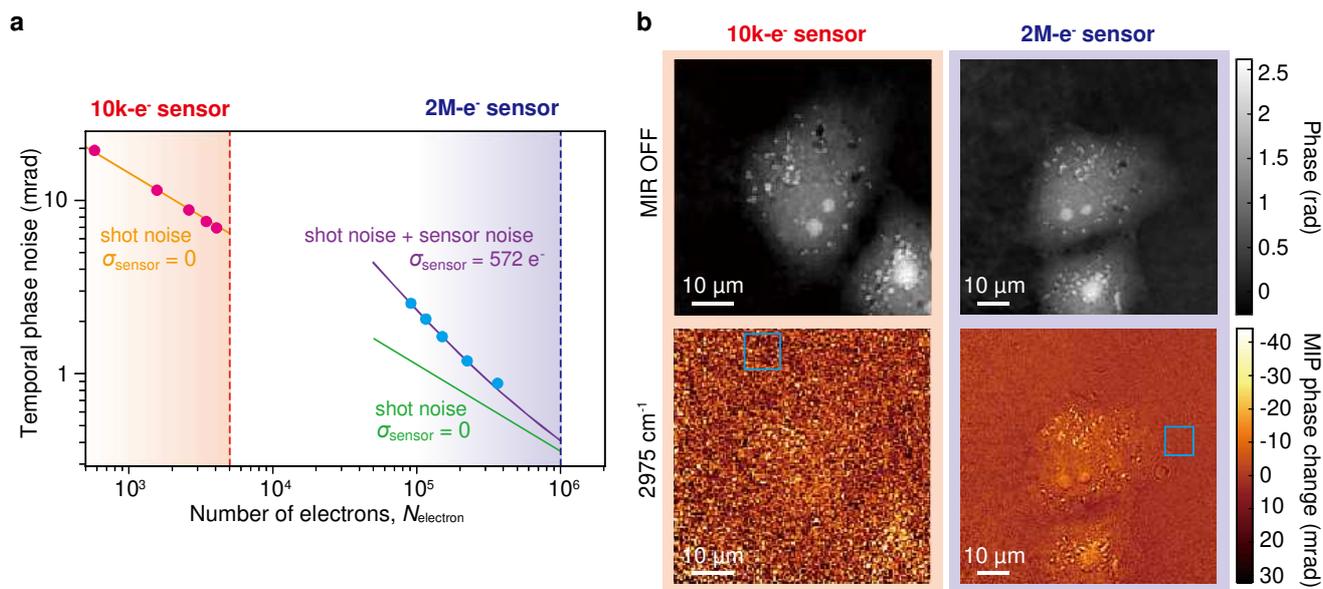

**Figure 4 Noise reduction in phase measurement with QPI**. **a** Dependence of temporal phase noise on the number of electrons generated in image sensors. Red and blue dots are measured data with a conventional 10k-e⁻ image sensor (acA2440-75um, Basler) on the left side and a high full-well-capacity 2M-e⁻ image sensor (Q-2HFW, Adimec) on the right side, respectively. Orange and green lines are theoretically estimated phase noise from Eq. 3 in Methods with sensor noise $\sigma_{sensor}$ = 0, and the purple line is that with $\sigma_{sensor}$ = 572 e⁻ (detailed parameters are written in Methods). **b** Comparison between MIP images taken with the two sensors: phase images (top) and MIP phase change images (bottom) of live COS7 cells taken with 10k-e⁻ (left) and 2M-e⁻ (right) sensors. The number of electrons per pixel, $N_{electron}$, is 2.82 × 10³ e⁻ and 2.34 × 10⁵ e⁻ with 10k-e⁻ and 2M-e⁻ sensors, respectively. MIR light at a wavenumber of 2,975 cm⁻¹ is irradiated on a spot size of 80 μm × 80 μm with a fluence of 1.1 nJ/μm².

**Video-rate MIP imaging of a single live cell.** We demonstrate MIP imaging of a live COS7 cell at 50 fps. Figure 5a is a phase image measured in MIR-OFF state, and 5b-d are MIP phase change images without water background subtraction, excited at 2,925, 2,964, and 3,188 cm⁻¹ MIR wavenumbers, respectively. Note that they are all single-frame images without averaging. The MIR pulse energy at the sample plane is ~6.5 μJ with a spot size of 87 μm × 87 μm. The image at 2,925 cm⁻¹ contains strong signals mainly from $CH_2$ bonds of lipid droplets, while the image at 2964 cm⁻¹ shows weaker signals from the lipid droplets and relatively higher contrasts from other areas such as the nucleus, which can be attributed to $CH_3$ bonds. The image at 3,188 cm⁻¹ shows signals that hardly reflect intracellular structures, mainly derived from OH bonds. Thus, different contrasts are observed at different MIR wavenumbers at an unprecedentedly high measurement rate of 50 Hz (20 ms measurement time per image). The phase noise, i.e., the spatial STD of 20 pixels × 20 pixels inside the blue box in Fig. 5b, is evaluated as 1.6 mrad with water background subtraction by following the procedure used in Fig. 4. Since the signal from the lipid droplets is ~100 mrad, the SNR is 63 (89 without water background subtraction). Hence, high-SNR MIP imaging beyond video rate is achieved for the first time.

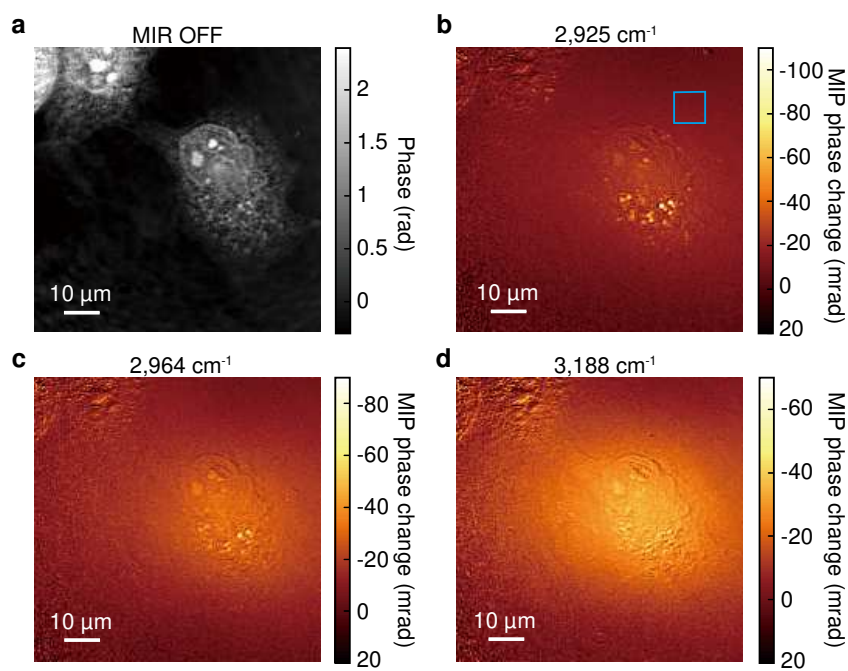

**Figure 5 Video-rate MIP imaging of a live COS7 cell. a** A phase image. **b-d** Single-frame MIP phase change images excited at 2,925, 2,964 and 3,188 cm$^{-1}$ with a spot size of 87 μm x 87 μm. The images are taken at a rate of 50 Hz (20 ms measurement time per image).

**Broadband MIP spectro-imaging of a single live cell.** We measure spectra of a single live COS7 cell and perform multivariate analysis as one of the applications utilizing the high SNR of our system. By scanning the wavenumber of the MIR light, 40 MIP phase change images are acquired in the range of 2,800~3,250 cm$^{-1}$. The MIR pulse energy at the sample plane is ~6.5 μJ with a spot size of 85 μm x 85 μm. The total acquisition time is ~10 min for 500 MIP phase change images averaged at each wavenumber. The hyperspectral data are subjected to multivariate analysis (Multivariate curve resolution, MCR)[34] to extract characteristic components (see Methods for details). Three clearly interpretable components are chosen for the analysis.

Figures 6a-c, d, e, and f show spatial distributions of the three MCR components, a phase image in a MIR-OFF state, a merged image of the three MCR images, and spectra of the three MCR components, respectively. In MCR1, the MIP contrasts are localized at extranuclear lipid droplets, and two peaks at 2,854 cm$^{-1}$ and 2,925 cm$^{-1}$ corresponding to symmetric and asymmetric stretching vibrations of CH$_2$ bonds appear in the spectrum, indicating that MCR1 mainly consists of lipids. In MCR2, the MIP contrasts are localized at the nucleus and nucleolus, and the spectrum is heavily influenced by the peak attributed to CH$_3$, indicating that it is a component with equal contributions of CH$_2$ and CH$_3$ bonds, which can be mainly attributed to proteins. In MCR3, uniformly distributed contrasts outside the cell can be recognized, and its spectrum resembles that of OH bonds, which have an absorption peak around 3,400 cm$^{-1}$ and monotonically increasing absorption towards higher wavenumbers in the observed wavenumber region[17], indicating that water is the main contributor. Thus, based on the vibrational modes of CH$_2$, CH$_3$, and OH, we are able to separate the three basic components of the cell. The spectral shapes are slightly unnatural in the sense that the high

wavenumber side of MCR1 is elevated, and the CH$_3$ absorption peak of MCR2 is ambiguous. However, the problem does not occur in a similar measurement where the medium is replaced with D$_2$O-based phosphate-buffered saline (PBS) to eliminate the effect of absorption by OH bonds (see Supplementary Note3 and Fig. S2). Hence, it can be considered that the spectra are slightly distorted during MCR analysis due to the stronger absorption of water compared to other components. This can be solved by expanding the measurement spectral region.

The MCR1 component at 2,925 cm$^{-1}$ (CH$_2$ peak) induces the MIP phase changes of 40 mrad at the lipid droplets, while the MCR2 component at 2,945 cm$^{-1}$ (CH$_3$ peak) induces only 14 mrad changes in the nucleolus, and 3 mrad changes in the cytoplasm, which is calculated with the procedure shown in Methods. Since the phase noise of our QPI is ~1.1 mrad, even smaller phase changes can be detected without averaging, although 500 images are averaged at each wavenumber in this work. This is because the current system requires a long time to save the measured data and also to scan the MIR wavenumber. Therefore, the total acquisition time stays the same with an average of 500 images. This situation can be solved by speeding up the controlling system. The estimated maximum temperature rise of this measurement is ~8 K for a lipid droplet (a sphere with a diameter of 3 µm) and ~2 K for a nucleolus (a sphere with a diameter of 5 µm), which quickly decays within ~2 and ~7 µs, respectively. These amounts of transient temperature rise have been proven to be safe for live cells[35].

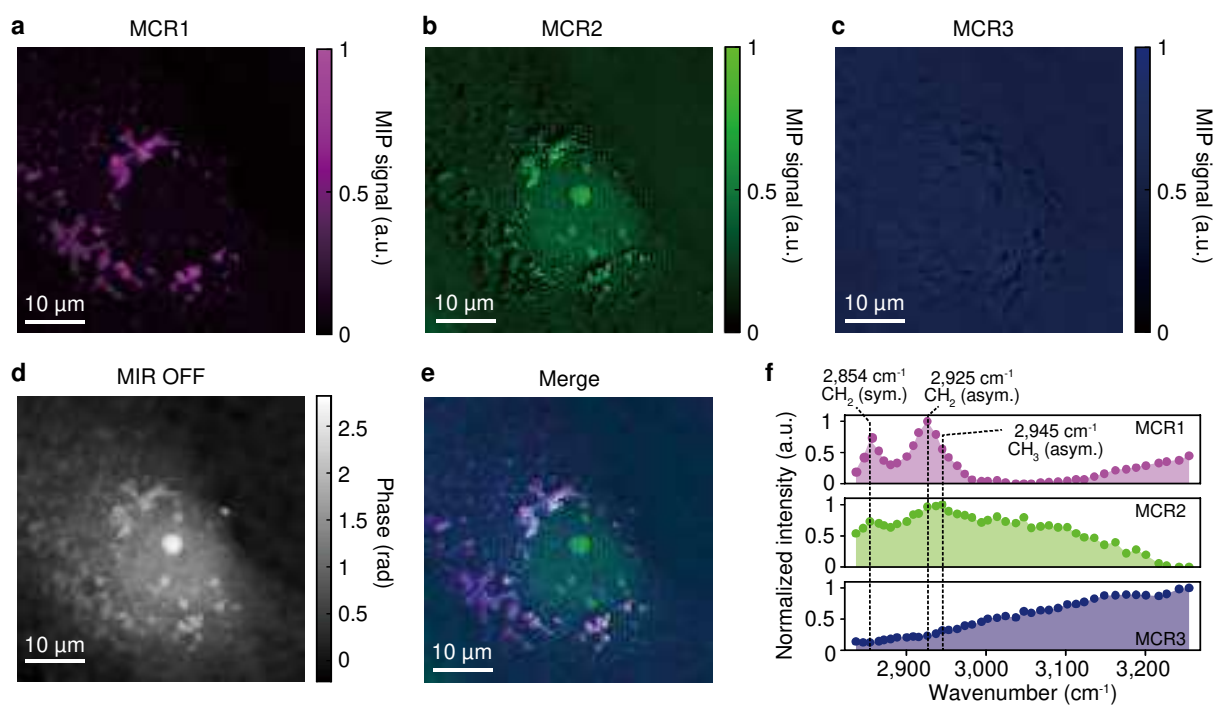

**Figure 6 Multivariate (MCR) analysis of a MIP spectro-image of a live COS7 cell.** MIP phase change images excited at 40 different wavenumbers in the range of 2,800~3,250 cm$^{-1}$ are analyzed by the MCR method. **a-c** Images of each MCR component. **d** A phase image. **e** A merged image of the three MCR images. **f** MIR spectra of each MCR component. MCR: multivariate curve resolution. sym.: symmetric vibrations. asym.: asymmetric vibrations.

**Discussions**

We make an SNR comparison between our system and the previous state-of-the-art wide-field MIP imaging system based on QPI[21]. The MIR pulse energy of our system is 6.5 µJ, while that of the previous work is 110 nJ, enabling ~59-times higher MIP phase change generation with our system. For visible imaging, the sensor's full-well capacity of our system is 2 Me$^-$, while that of the previous work is 30 ke$^-$. Considering the sensor and shot noises, ~7.1-times higher SNR is achievable with our system. In the current system, however, we only use $N_{electron}$=360 ke$^-$ due to limitation of the photon budget that can be coupled to the single-mode optical fiber without damage, which gives ~3.5-times higher SNR if we assume the previous work fully uses the sensor's capacity ($N_{electron}$=15 ke$^-$). In total, our system can provide a higher SNR than the previous work by up to ~420 times if using the full capability of the system or by ~210 times with the current demonstration with the limited visible photon budget. We note that this comparison is based upon the same conditions with the pulse repetition rate of 1 kHz and the sensor's frame rate of 100 Hz.

There is room for further technical improvements in our microscope. The first is to broaden the tunable spectral range of the MIR nanosecond OPO, covering the molecular fingerprint region by using other nonlinear crystals such as AGS[36], BGSe[37], or OP-GaP. This could enable ultra-broadband MIR spectroscopic imaging in the range of 600-3,700 cm$^{-1}$[35]. The second is to further improve the detection sensitivity and imaging speed. In this experiment, the sensor's full-well capacity (2 Me$^-$) and maximum frame rate (500 fps) are not used to the full extent due to insufficient light intensity. This is because the intensity of irradiated visible light is limited by the damage threshold of the single-mode fiber. This can be solved by using a large-core single-mode fiber often used for high-power lasers. The intensity of visible light can be increased by a factor of ~20, which enables ultrafast MIP imaging at a maximum imaging rate of 250 fps (limited by the frame rate of our image sensor) with an improved SNR up to 250. It could also be possible to further increase the detection sensitivity by combining our system with a highly sensitive QPI using a wavefront shaping technique (ADRIFT-QPI)[24]. The third is an extension to high-speed 3D imaging. QPI can be extended to optical diffraction tomography (ODT)[38], in which 3D refractive index distributions can be obtained by imaging with, e.g., multiple illuminations at different angles. With a commercially available high-speed spatial light modulator (SLM), which can change the illumination pattern at ~500 Hz, it could be possible to perform the world's first 3D vibrational imaging at a video rate that has not yet been achieved even with coherent Raman imaging.

Finally, we discuss the consequences when the above improvements are realized. Currently, only a spectral bandwidth of ~450 cm$^{-1}$ in the high wavenumber region is available with a high pulse energy of around 10 µJ. Therefore, the distinguishable molecular species are limited to those containing $CH_2$, $CH_3$, and OH bonds. When ultra-broadband MIR spectroscopic imaging is achieved, molecules such as DNA, RNA, multiple types of proteins, lipids, etc., could be distinguished by multivariate analyses. Since our microscope can obtain information on the dry mass density of total intracellular biomolecules from quantitative phase images[39], correlation analysis with MIP images could enable molecule-specific mass density imaging[40]. Other promising applications with high-speed imaging are, e.g., video-rate observations of sub-second biological phenomena such as cell signaling[41], bacterial spore germination[42], or high-speed infrared hyperspectral image acquisition.

## Methods

**Preparation of biological samples.** COS7 cells are cultured on a $CaF_2$ substrate with a thickness of 500 µm in high glucose Dulbecco's modified eagle medium with L-glutamine, phenol red, and HEPES (FUJIFILM Wako) supplemented with 10% fetal bovine serum (Cosmo Bio) and 1% penicillin-streptomycin-L-glutamine solution (FUJIFILM Wako) at 37 °C in 5% $CO_2$, and are sandwiched with another CaF2 substrate before imaging. For live-cell imaging in $D_2O$ environment (Fig. S2), the medium is replaced by $D_2O$-based PBS.

**Thermal conduction simulations.** For Figs. 1a-d, the spherically symmetric 3-D thermal conduction equation is exploited. For Fig. 1f, Eq. 2 is used with the boundary conditions between water and $CaF_2$ substrates given by

$$K_{\text{water}} \left(\frac{\partial T_{\text{water}}(x,y,z,t)}{\partial z}\right)_{z=\text{boundary}} = K_{\text{CaF}_2} \left(\frac{\partial T_{\text{CaF}_2}(x,y,z,t)}{\partial z}\right)_{z=\text{boundary}},$$

where $K_{\text{water}}$ and $K_{\text{CaF}_2}$ are the thermal conductivities of water and $CaF_2$, respectively. In the calculations, the thermal diffusivities and thermal conductivities of water and $CaF_2$ substrates are 0.146 and 2.92 µm²/µs, and 0.618 and 9.71 W/m·K, respectively.

**Parameters in phase sensitivity evaluation.** The temporal phase noise, $\sigma_{\text{phase}}$, can be described as,

$$\sigma_{\text{phase}} = 2\sqrt{\frac{[N_{\text{electron}}+(\sigma_{\text{sensor}})^2]A_{\text{aperture}}}{v^2(N_{\text{electron}})^2 A_{\text{sensor}}}}, \quad \text{(Eq.3)}$$

where $\sigma_{\text{sensor}}$ denotes the sensor noise, $v$ the visibility of the hologram, $A_{\text{sensor}}$ and $A_{\text{aperture}}$ the numbers of pixels in total and cropped areas in the spatial frequency space. The number of electrons contributing to the reconstruction of a phase image, $N_{\text{electron}}$, is calculated from the image sensor output value with sensor's parameters of full-well capacity, bit depth (2M-e⁻ sensor: 11 bit, 10k-e⁻ sensor: 16 bit), and gain (2M-e⁻ sensor: 1.73, 10k-e⁻ sensor: 1). To obtain $\sigma_{\text{sensor}}$, a series of images are taken without light, and the temporal standard deviation of the difference images between adjacent frames is calculated, which is converted to the number of electrons. The visibility $v$ is evaluated by the procedure described in Supplementary Note 2. The numbers of pixels $A_{\text{sensor}}$ and $A_{\text{aperture}}$ are 2,073,600 (1,440 pixels x 1,440 pixels) and 47,144 ( π /4 x 245 pixels x 245 pixels) for the 2M-e⁻ sensor, and 1,046,529 (1,023 pixels x 1,023 pixels) and 31,731 (π/4 x 201 pixels x 201 pixels) for the 10k-e⁻ sensor, respectively.

**MCR analysis.** Prior to MCR analysis, the spatial MIP phase change contrasts reflecting the MIR beam profile are corrected by dividing the MIP phase change images of cells by normalized MIP phase change images of water without cells. Also, the wavenumber-dependent power variation of the MIR light is normalized with the data shown in Fig. 3a. MCR analysis is performed using pyMCR developed by NIST with a non-negativity constrained least-squares regressor. The spectral data with water background subtraction at the nucleolus and lipid droplets and those without water background subtraction outside the cell are used for the initial input spectra in MCR.

We calculate the MIP phase change contributed by each MCR component by the following procedure. MCR decomposes the hyperspectral data into matrices of the concentration distribution $\boldsymbol{C}$ and the pure spectra $\boldsymbol{S}$ for each MCR component $i$,

$$H(x,k) = \boldsymbol{CS} = \sum_i C_i(x)S_i(k),$$

where $x$ and $k$ denote the location and the MIR wavenumber, respectively. Each component's contribution to the MIP phase change at (x, k) can be calculated as

$$R_i(x,k) = \frac{C_i(x)S_i(k)}{\sum_i C_i(x)S_i(k)} \times \Delta\emptyset(x,k),$$

where $\Delta\emptyset(x,k)$ is the raw MIP phase change at (x, k).

## Data availability

The data provided in the manuscript is available from T.I. upon request.

## Acknowledgments

This work was financially supported by Japan Society for the Promotion of Science (20H00125), JST PRESTO (JPMJPR17G2), Precise Measurement Technology Promotion Foundation, Research Foundation for Opto-Science and Technology, Nakatani Foundation, and UTEC-UTokyo FSI Research Grant. We thank Masaki Yumoto for his advice about nanosecond MIR lasers and Akira Kamijo and Kohki Horie for their manuscript review.

## Author contributions

M.T and V.R.B designed and constructed the optical systems. H.S wrote a program to control the systems. G.I and K.T performed the experiments and analyzed the data. T.I. supervised the entire work. G.I., K.T., and T.I. wrote the manuscript with inputs from the other authors.

## Competing interests

K.T., M.T., and T.I. are inventors of patents related to MIP-QPI.

# Supplementary Information for
# Mid-infrared photothermal single-live-cell imaging beyond video rate

Genki Ishigane[1+], Keiichiro Toda[1+], Miu Tamamitsu[1,2], Hiroyuki Shimada[2], Venkata Ramaiah Badarla[2], and Takuro Ideguchi[1,2,*]

[1] Department of Physics, The University of Tokyo, Tokyo, Japan

[2] Institute for Photon Science and Technology, The University of Tokyo, Tokyo, Japan

[+] These authors contributed equally to this work

[*] Corresponding author: ideguchi@ipst.s.u-tokyo.ac.jp

**Supplementary Note 1: Comparison of parameters of the demonstrated wide-field MIP imaging systems**

Table S1 shows the various parameters of MIR light, visible light, image sensors, and imaging frames of the previously demonstrated wide-field MIP imaging systems and our developed system in this work.

**Table S1 Parameters of previously demonstrated wide-filed MIP imaging systems and our system.**

| Ref. (Year) | MIR light<br>Pulse energy (excitation area)<br>Pulse duration, repetition rate<br>Wavenumber tuning range | Visible light<br>Pulse duration<br>Repetition rate | Image sensor<br>Frame rate<br>Full-well capacity | Imaging frame rate |
|---|---|---|---|---|
| [1] (2019) | Pulsed OPO<br>~250 nJ* (~40 μm x 40 μm)<br>50 ns*, 20 kHz<br>1,175-1,800 cm$^{-1}$ | LED<br>914 ns<br>20 kHz | 2.5 kHz<br>19 ke$^-$ | 2 Hz |
| [2] (2019) | CW QCL<br>16-40 mW (~40 μm x 60 μm)*<br>2 ms, 250 Hz<br>1,450-1,640 cm$^{-1}$ | LED<br>CW | 10 kHz<br>~20 ke$^-$* | 10 Hz |
| [3] (2019) | Pulsed OPO<br>~110 nJ (~40 μm x 60 μm)*<br>10 ns, 1 kHz<br>2,700-3,600 cm$^{-1}$ | SHG of fs laser<br>900 ns<br>1 kHz | 100 Hz<br>30 ke$^-$ | 1 Hz |
| [4] (2020) | Pulsed QCL<br>~100 nJ (~40 μm x 40 μm)*<br>1 μs, 1 kHz<br>1,450-1,640 cm$^{-1}$ | SHG of ns laser<br>~10 ns<br>1 kHz | 100 Hz<br>10 ke$^-$* | 0.02 Hz |
| [5] (2020) | CW QCL<br>500 mW* (~460 μm x 460 μm)<br>0.83 ms, 600 Hz<br>900-1,900 cm$^{-1}$ | LED<br>CW | 500 Hz<br>~1 Me$^-$ | ~0.1 Hz |
| [6] (2021) | Pulsed QCL<br>~100 nJ (~30 μm x 70 μm)<br>1 μs, 1 kHz<br>1,450-1,640 cm$^{-1}$ | SHG of ns laser<br>~10 ns<br>1 kHz | 20 Hz<br>10 ke$^-$ | ~0.1 Hz* |
| [7] (2021) | Pulsed QCL<br>~40 nJ* (~70 μm x 70 μm)<br>1 μs, 200 kHz<br>1,500-1,700 cm$^{-1}$ | SHG of fs laser<br>200 ns<br>200 kHz | 400 Hz | 0.2 Hz |
| This work (2022) | Pulsed OPO<br>~10 uJ (~85 μm x 85 μm)<br>10 ns, 1 kHz<br>2,800-3,250 cm$^{-1}$ | SHG of ns laser<br>10 ns<br>1 kHz | 100 Hz<br>2 Me$^-$ | 50 Hz |

*These values are estimated by the authors from the displayed data and specification sheets of devices because these are not clearly specified in the referenced papers.

**Supplementary Note 2: Procedures of phase image reconstruction and visibility evaluation**

Figure S1a is a hologram captured by the image sensor. Figure S1b is the Fourier transform of Fig. S1a, including three spectral components, a non-interferometric term in the center, and two neighboring interferometric terms. Inverse Fourier transform of the cropped area (within the circle of Fig. S1b) provides a complex amplitude image of the electric field, from which the quantitative phase image can be obtained by taking the phase components (see Fig. S1e). Figure S1f shows the visibility $v$ $(= 2\beta/\alpha)$, where $\alpha$ and $\beta$ are the amplitudes of the inverse Fourier transforms of the interferometric and non-interferometric terms (Fig. S2c and 2d), respectively. The number of electrons contributed to the reconstruction of the phase image $(=N_{electron})$ is calculated from $\beta$ with the sensor's parameters such as full-well capacity, bit depth, and gain.

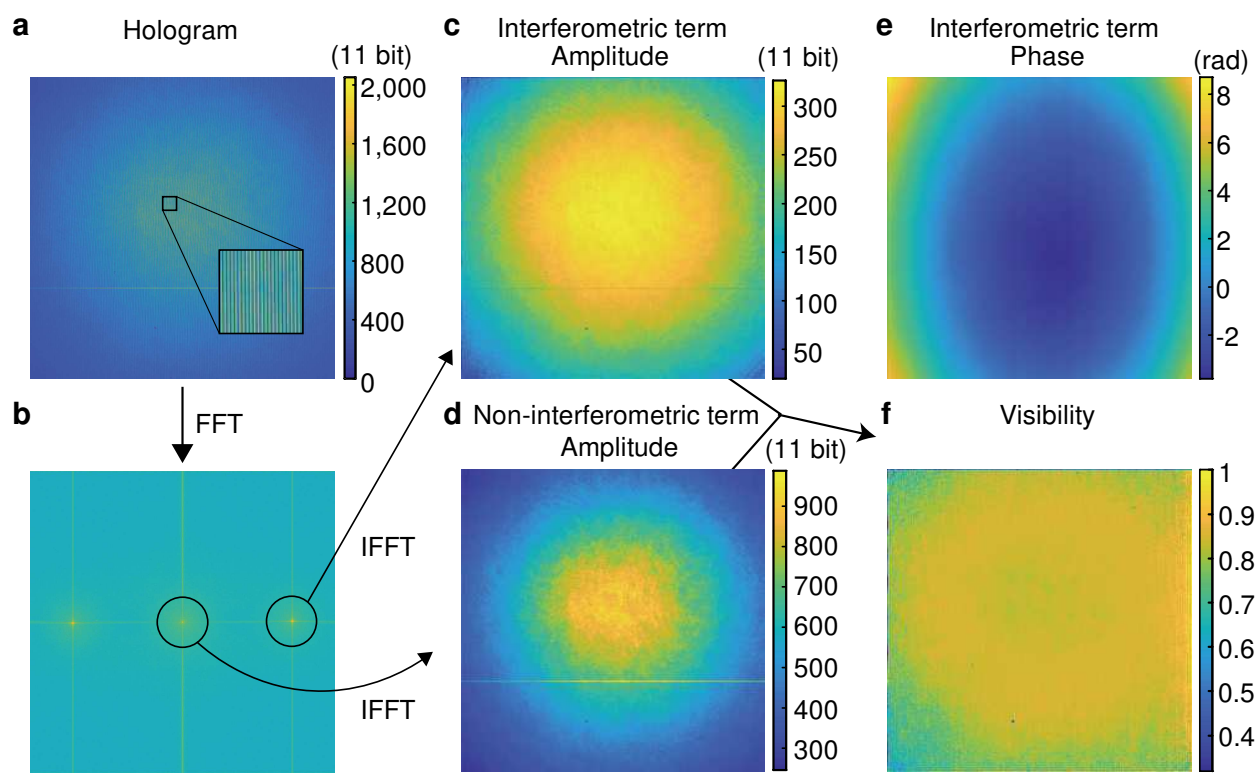

**Figure S1 Procedure to calculate the phase image and the visibility from a measured hologram. a** Measured hologram. **b** Fourier transform of **a**. **c, e** Amplitude and phase images derived from inverse Fourier transform of the interferometric term in **b**. **d** Amplitude image derived from inverse Fourier transform of the non-interferometric term in **b**. **f** Visibility calculated from **c** and **d**.

**Supplementary Note 3: Broadband MIP spectro-imaging of a single living cell in D$_2$O**

We measure spectra of a single live COS7 cell in D$_2$O-based PBS to eliminate the effect of absorption by OH bonds. 40 MIP images are acquired in the range of 2,800~3,250 cm$^{-1}$ (under the same conditions as Fig. 6) and subjected to multivariate analysis. Figures S2a-c, d, e and f show spatial distributions of the three MCR components, a phase image in a MIR-OFF state, a merged image of the three MCR images, and spectra of the three MCR components, respectively.

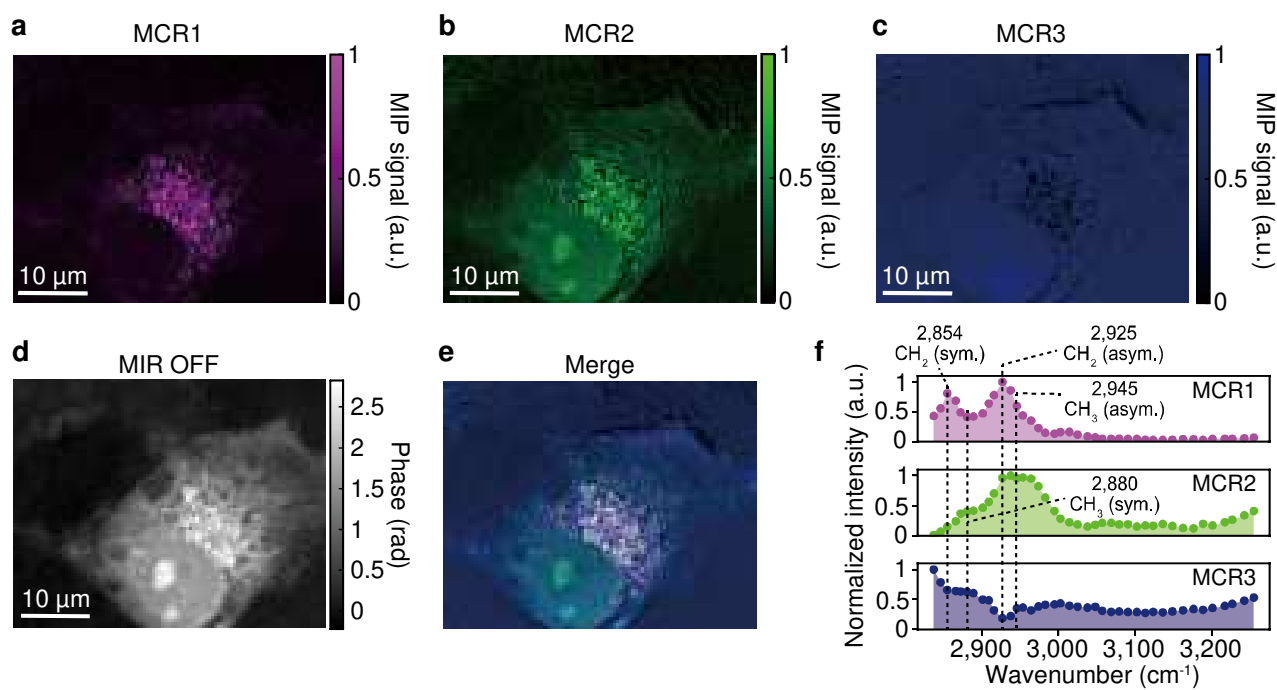

**Figure S2 Multivariate analysis of living COS7 cells in D$_2$O-based PBS.** 40 MIP images are acquired in the range of 2,800~3,250 cm$^{-1}$. **a-c** Images of each MCR component. **d** A phase image. **e** A merged image of the three MCR images. **f** MIR spectra of each MCR component. MCR: multivariate curve resolution. sym.: symmetric vibrations. asym.: asymmetric vibrations.